\documentclass[fleqn,12pt,twoside]{article}
\usepackage{espcrc1}

\usepackage{graphicx}
\usepackage[figuresright]{rotating}

\newcommand{\AmS}{{\protect\the\textfont2
  A\kern-.1667em\lower.5ex\hbox{M}\kern-.125emS}}
\renewcommand{\vec}[1]{\mbox{\boldmath $#1$}}

\title{Fusion and breakup in the reactions of $^{6,7}$Li and $^9$Be}

\author{K. Hagino\address{Yukawa Institute for Theoretical Physics,
Kyoto University, Kyoto 606-8502, Japan}, 
M. Dasgupta\address{Department of Nuclear Physics, Research School of Physical 
Sciences and Engineering, Australian National University, 
Canberra ACT 0200, Australia} and 
D.J. Hinde\addressmark}

\begin{document}

\maketitle

\begin{abstract}
We develop a three body classical trajectory Monte Carlo (CTMC) method 
to dicsuss the effect of the breakup process on heavy-ion fusion 
reactions induced by weakly bound nuclei. This method follows the
classical trajectories of breakup fragments after the breakup takes
place, and thus provides an unambiguous separation between complete and
incomplete fusion cross sections. Applying this method to the fusion
reaction $^{6}$Li + $^{209}$Bi, we find that there is a significant 
contribution to the total complete fusion cross sections 
from the process where all the breakup fragments are
captured by the target nucleus (i.e., the breakup followed by complete
fusion). 
\end{abstract}

\section{Introduction}

The fusion reactions of neutron-rich nuclei around the Coulomb 
barrier provide a good opportunity to study the interplay between 
quantum tunneling and the breakup process.
An important question here is: does the breakup process influence 
fusion reactions in a similar way as the inelastic process, which leads
to subbarrier enhancement \cite{DHRS98} 
of fusion cross sections over predictions for a single barrier? 
In addressing this question, measurements with 
weakly bound stable nuclei, such as $^9$Be, $^6$Li, and $^7$Li, have been 
proved to be useful\cite{D99,D02,P02}. 
These beams are currently much more intense than 
radioactive beams, allowing more precise and extensive experimental 
studies. Also, these nuclei predominantly 
break up into charge fragments 
($^9$Be $\to$ 2$\alpha +n$ or $\alpha + ^5$He, 
$^7$Li $\to$ $\alpha + t$, and $^6$Li $\to$ $\alpha + d$),  
which are 
more easily detected. This allows a clean separation of the products 
of complete fusion from those of  
incomplete fusion, where only a part of the projectile is captured. 

To describe the effect of breakup on
fusion reactions theoretically, a model has to take into account 
the following three different processes: (i) the projectile as a whole
is captured by the target, (ii) only one of the breakup fragments is
captured, (iii) all the breakup fragments are absorbed after the
breakup takes place near the target nucleus. Process (ii) is
referred to as 
an incomplete fusion reaction, while both of processes (i) and
(iii) lead to complete fusion. Calculations based on the
continuum-discretized coupled-channels method may account for the
total fusion cross section (the sum of all the processes) and/or the
separate contribution of process (i)\cite{HVDL00,DTT02}, but it is 
not easy to compute cross sections for process (iii), that is, 
the breakup followed by complete fusion. 
In order to model this process, 
we need to follow the trajectories of the breakup fragments to
determine whether one or both fragments are captured by the target
nucleus. In this contribution, we present such a model\cite{D02}, where 
many classical trajectories are 
sampled by the Monte Carlo method to 
compute the total complete fusion cross section. 
We note that a very similar model, called the three-body classical
trajectory Monte Carlo (CTMC) method, 
has been developed in the field of atomic and
molecular physics, and has been successfully applied to ion-atom
ionization and charge transfer collisions \cite{AP66,C83,FCRB00}. 

\section{Three-body classical trajectory Monte Carlo (CTMC) model}

Let us assume the following Hamiltonian for 
a reaction of a projectile nucleus 
which consists of two cluster fragments $P_1$ and $P_2$,
\begin{equation}
H=\frac{\vec{p}_1^2}{2m_1}+\frac{\vec{p}_2^2}{2m_2}
+\frac{\vec{p}_T^2}{2m_T}+V_{12}(\vec{r}_1-\vec{r}_2)
+V_{1T}(\vec{r}_{1}-\vec{r}_T)+V_{2T}(\vec{r}_{2}-\vec{r}_T),
\end{equation}
where $\vec{p}_1$, $\vec{p}_2$, and $\vec{p}_T$ are the momenta of 
fragment $P_1$,  $P_2$, and of the target, respectively. 
$\vec{r}_1$, $\vec{r}_2$, and $\vec{r}_T$ are the corresponding 
coordinates. The idea of the three-body classical trajectory model 
is to solve this Hamiltonian classically in the two-dimensional
($x,y$) plane \cite{D02}. 
We assume that at $t$=0 the target is at rest at the origin, while the 
center of mass of the projectile is at $(x,y)=(R_{\rm max},b)$ moving 
towards the $-x$ direction. 
Here, $R_{\rm max}$ is some large number, while $b$ is the 
impact parameter. 
The relative motion between the projectile fragments 
is confined at $t=0$ inside the potential $V_{12}$ at the energy
corresponding to
the breakup $Q$-value, $Q_{\rm bu}$, with a random orientation $\theta$
with
respect to the $x$ axis. 
The initial conditions thus read, 
\begin{eqnarray}
x_1(0)&=&R_{\rm max}-\frac{m_2}{m_1+m_2}\,r_0\,\cos\theta,~~~~~
y_1(0)=b+\frac{m_2}{m_1+m_2}\,r_0\,\sin\theta, \\
x_2(0)&=&R_{\rm max}+\frac{m_1}{m_1+m_2}\,r_0\,\cos\theta,~~~~~
y_2(0)=b-\frac{m_1}{m_1+m_2}\,r_0\,\sin\theta, \\
v_{x1}(0)&=&-v_{\rm lab}-\frac{1}{2m_1}p_{\rm rel}\cos\theta, 
~~~~~~~~v_{y1}(0)=\frac{1}{2m_1}p_{\rm rel}\sin\theta, \\
v_{x2}(0)&=&-v_{\rm lab}+\frac{1}{2m_2}p_{\rm rel}\cos\theta, 
~~~~~~~~v_{y2}(0)=-\frac{1}{2m_2}p_{\rm rel}\sin\theta, \\
x_T(0)&=&y_T(0)=v_{xT}(0)=v_{yT}(0)=0,
\end{eqnarray}
where 
\begin{equation}
v_{\rm lab}=\sqrt{\frac{2}{m_1+m_2}[E_{\rm lab}-V_{1T}(r_{1T}(0))
-V_{2T}(r_{2T}(0))]},
\end{equation}
is the initial velocity for the 
center of mass of the projectile, and 
$p_{\rm rel}=\sqrt{2\mu[Q_{\rm bu}-V_{12}(r_{12}(0))]}$ with 
$\mu=m_1m_2/(m_1+m_2)$. The value of $r_0$ is restricted to 
$0 \leq r_0 \leq r_t$, where $r_t$ satisfies $V_{12}(r_t)=Q_{\rm bu}$. 

With these initial conditions, we solve the Newtonian equations to 
follow the time evolution of the trajectories for each particle, 
$x_i(t), y_i(t), v_{xi}(t)$, and $v_{yi}(t), (i=1,2,T)$. 
At each time, we monitor the relative distance between the 
particles. We assume that the projectile fragment $P_i~(i=1,2)$ is 
absorbed by the target nucleus when the relative distance between 
the fragment and the target is smaller than 
$r_{iT} < 1.1 (A_i^{1/3} + A_T^{1/3})$ fm, and that the breakup 
occurs when the relative distance between $P_1$ and $P_2$ is larger
than the barrier radius for the potential $V_{12}$. 
By sampling many
trajectories 
for random values of $r_0$ and $\theta$ 
with the Monte Carlo method, we calculate partial 
cross sections for a given impact parameter $b$. Total cross sections 
are then obtained by integrating them over the value of $b$.

\begin{figure}[htb]
\begin{minipage}[b]{75mm}
\includegraphics[scale=0.45]{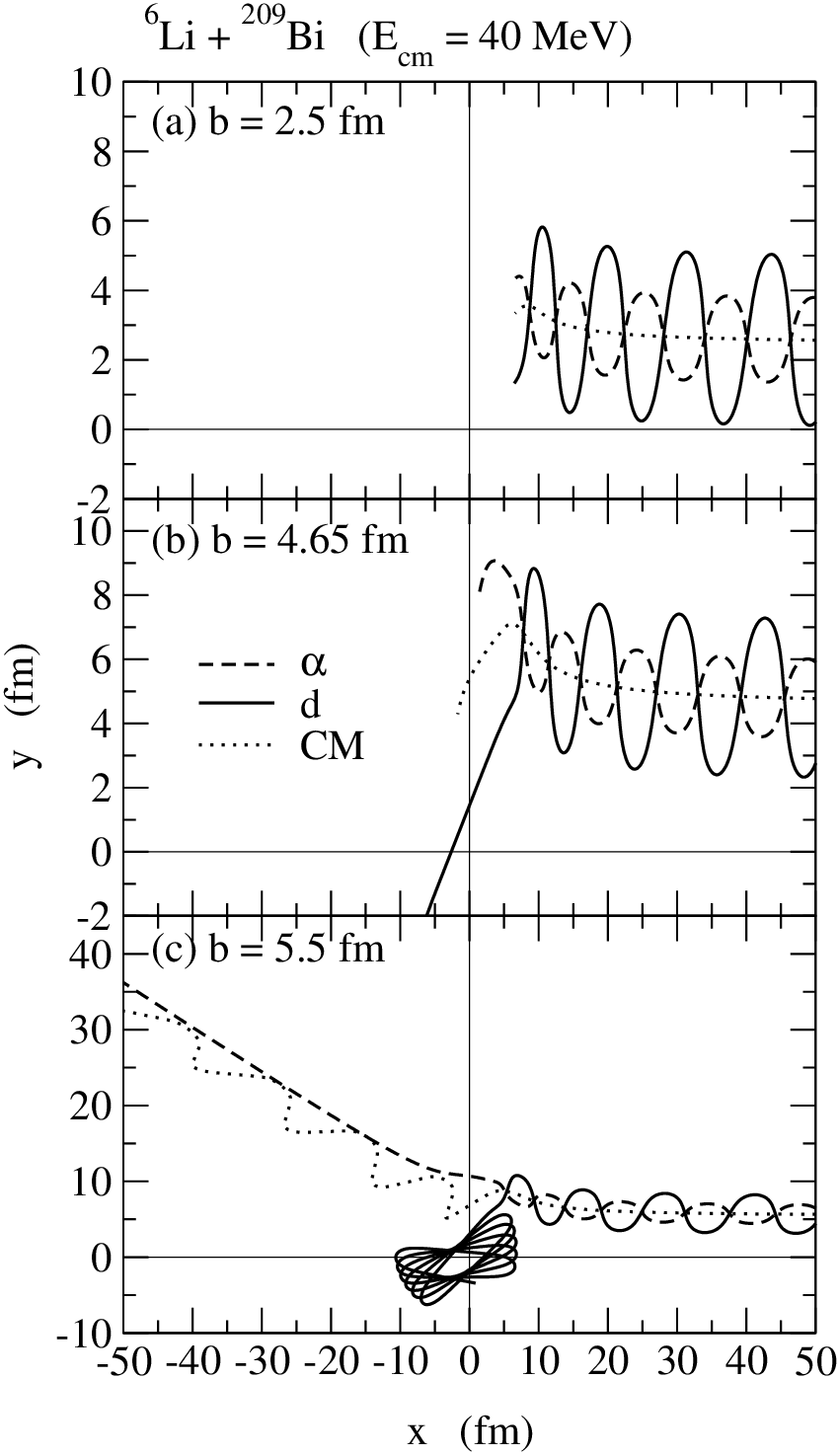}
\caption{Examples of classical trajectories 
for the $^6$Li+$^{209}$Bi reaction. 
Fig. 1(a), (b), and (c) are for the impact parameter $b$=2.5, 
4.65, and 5.5 fm at $E_{cm}$=40 MeV, respectively. 
The dashed and the solid lines correspond to the trajectories for 
the $\alpha$ particle and deuteron, respectively, while the 
dotted line is for the center of mass of the projectile fragments.}
\end{minipage}
\hspace{\fill}
\begin{minipage}[b]{75mm}
\includegraphics[width=6cm,clip]{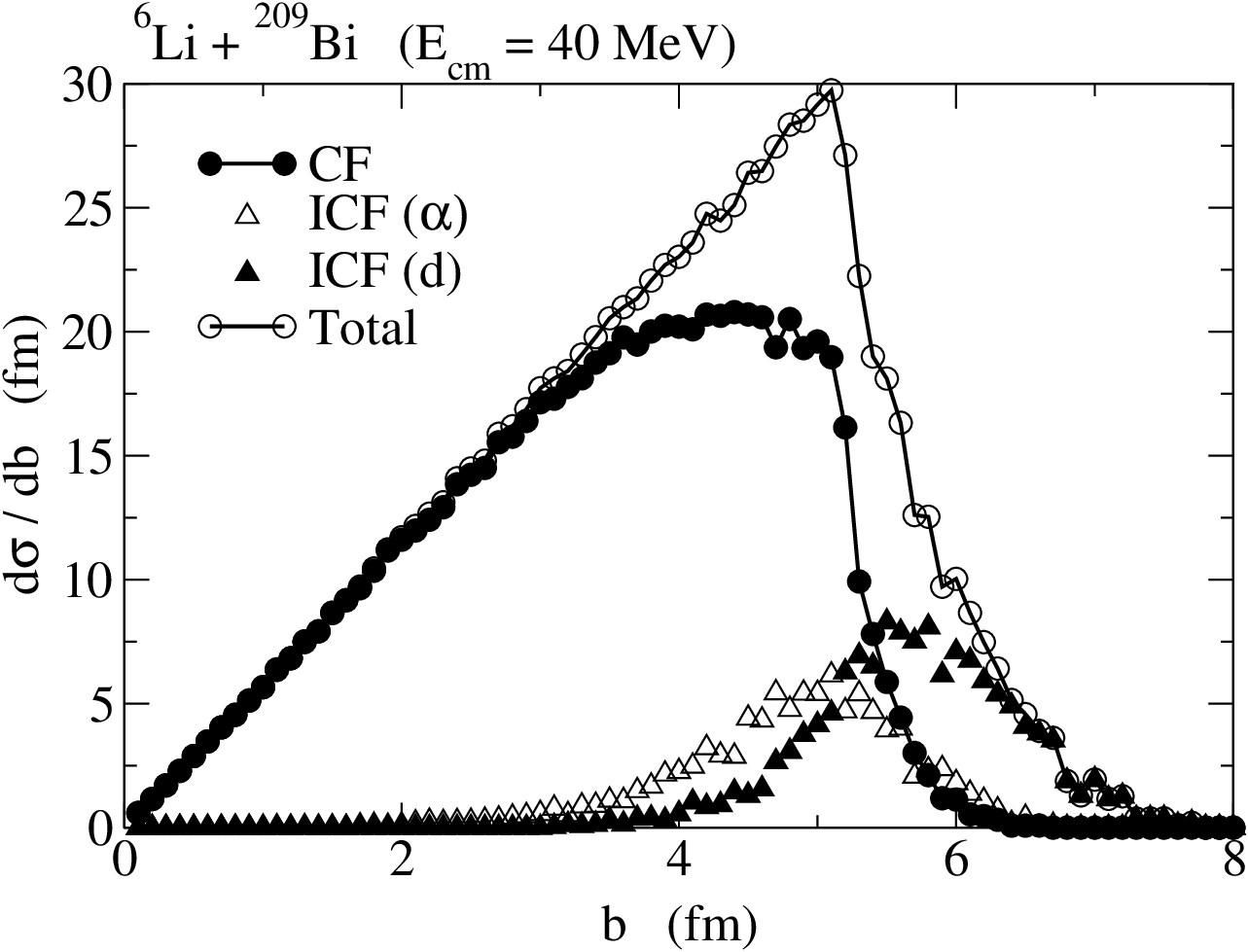}
\caption{The partial fusion cross sections $d\sigma/db$ for the 
$^6$Li+$^{209}$Bi reaction at $E_{cm}$=40 MeV obtained with the 
three-body classical trajectory Monte Carlo method. }
\vspace*{0.5cm}
\includegraphics[width=6cm,clip]{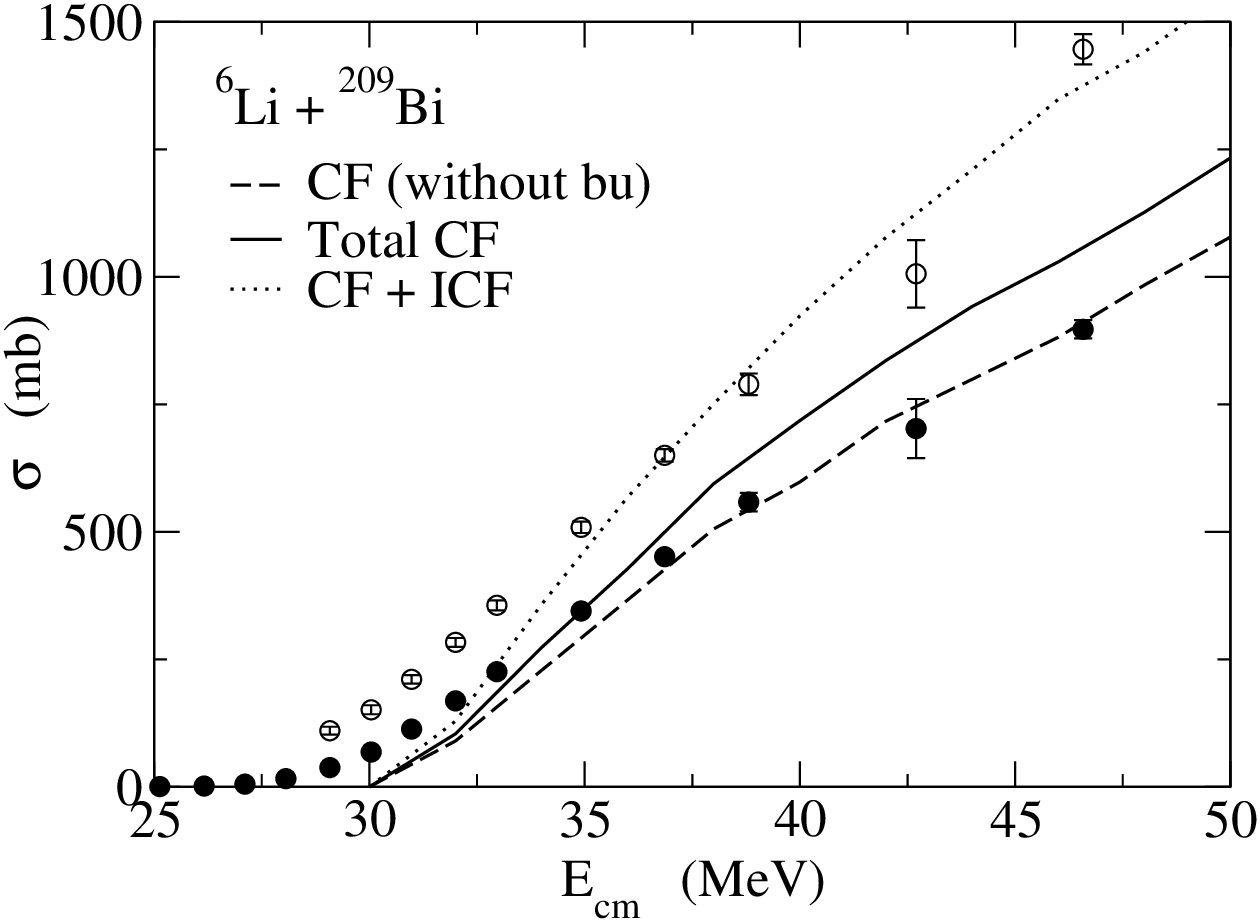}
\caption{The complete fusion (CF) and the incomplete fusion (ICF) 
cross sections for the $^6$Li+$^{209}$Bi reaction. The 
experimental data are taken from Ref. \cite{D02}. }
\end{minipage}
\end{figure}

\section{Application to $^6$Li + $^{209}$Bi reaction}

We now apply the three-body classical trajectory Monte Carlo method 
to the $^6$Li + $^{209}$Bi system and discuss the effect of breakup 
of $^6$Li into $d+\alpha$ on the fusion reaction near the Coulomb
barrier. We assume the Woods-Saxon form in calculating 
$V_{12}, V_{1T}$, and $V_{2T}$, using the same values for the
parameters as in Refs. \cite{TN81,GSW70}. We take $R_{\rm max}$ = 200
fm. 
Fig. 1(a) -- 1(c) shows sample trajectories at $E_{\rm cm}$ = 40 MeV. 
These are for $b$=2.5, 4.65, and 5.5 fm, respectively. For relatively
small values of the impact parameter (Fig. 1(a)), 
the projectile nucleus $^6$Li 
is captured by the target before breaking up into 
$\alpha$ and $d$, 
leading to complete fusion. For
$b$=5.5 fm (Fig. 1(c)), only the deuteron is captured while
the alpha particle escapes from the target. This is an example of 
incomplete
fusion. For intermediate values of $b$, breakup
takes place near the target, but both of the breakup fragments get 
absorbed (Fig. 1(b)). This is breakup followed by complete
fusion. 
Figure 2 shows the partial fusion cross sections 
for these processes as a function of the impact parameter, for 
$E_{cm}$=40 MeV. 

Figure 3 shows fusion cross sections as a function of energy in the 
center of mass frame. 
The dotted line denotes the sum of complete and incomplete fusion 
cross sections. 
The dashed line indicates the absorption of the whole projectile from the
bound states, while the solid line shows the total complete fusion cross
sections. The difference between the solid and the dashed line
thus represents the contribution from breakup followed by 
fusion of both fragments. This makes 
a significant
contribution to complete fusion at energies above the barrier. 

\section{Summary}

We present 
a three body classical trajectory Monte Carlo method for fusion 
of weakly bound nuclei. Applying this method to 
the 
$^6$Li + $^{209}$Bi reaction, 
we have demonstrated that there is a significant contribution of breakup 
followed by complete fusion to the total complete fusion cross 
sections. 
This process has not been considered in any coupled-channels
calculations so far, and the present result suggests that 
a consistent definition of complete fusion is necessary when one
compares experimental data with theoretical calculations.

\end{document}